\documentclass[aps,preprint,tightenlines]{revtex4}%
\usepackage{amsfonts}
\usepackage{amsmath}
\usepackage{amssymb}
\usepackage{graphicx}%
\begin{document}
\title[Casimir Energies in One Dimension]{Casimir Forces and Boundary Conditions in One Dimension: Attraction,
Repulsion, Planck Spectrum, and Entropy}
\author{Timothy H. Boyer}
\affiliation{Department of Physics, City College of the City University of New York, New
York, New York 10031}
\pacs{PACS number}

\begin{abstract}
Quantities associated with Casimir forces are calculated in a model wave
system of one spatial dimension where the physical ideas are transparent and
the calculations allow easy numerical evaluation.\ The calculations show
strong dependence upon fixed- or free-end (Dirichlet or Neumann) boundary
conditions for waves on a one-dimensional string, analogous to
infinitely-conducting or infinitely-permeable materials for electromagnetic
waves. \ 1) Due to zero-point fluctuations, a partition in a one-dimensional
box is found to be attracted to the walls if the wave boundary conditions are
alike for the partition and the walls, but is repelled if the conditions are
different. \ 2) The use of Casimir energies in the presence of zero-point
radiation introduces a natural maximum-entropy principle which is satisfied
only by the Planck spectrum for both like and unlike boundary conditions
between the box and partition. \ 3)The one-dimensional Casimir forces increase
or decrease with increasing temperature depending upon like or unlike boundary
conditions. The Casimir forces are attractive and increasing with temperature
for like boundary conditions, but are repusive and decreasing with temperature
for unlike boundary conditions. \ 4) In the high-temperature limit, there is a
temperature-independent Casimir entropy for like boundary conditions, but a
vanishing Casimir entropy for unlike boundary conditions. \ These
one-dimensional results have counterparts for electromagnetic Casimir forces
in three dimensions.

\end{abstract}
\date{October 24, 2002}
\startpage{1}
\endpage{2}
\maketitle

\section{Introduction}

There has been an ever-increasing interest in Casimir forces during the past
half-century.[1] \ What began as an obscure phenomenon in electrodynamics[2]
has became of general theoretical interest in quantum field theory[3], has
been connected to wave phenomena in acoustics[4] and boating,[5] and is now of
technological concern in connection with microelectromechanical devices.[6]
\ In particular, there have been accurate experimental measurements of
electromagnetic Casimir forces[7] and suggestions that such forces will create
a "stiction" which will bind surfaces together at small separations. \ In
order to overcome such stiction, the repulsive aspects of Casimir forces have
recently been reinvestigated.[8] \ In this article we review some aspects of
thermodynamic Casimir forces and energies in a simple model system allowing
extraction of the essential physics with a minimum of mathematical complexity.

Our work presents simple examples of physical aspects of Casimir forces in one
spatial dimension. \ We begin with an example of positive and negative Casimir
energies at zero-temperature in connection with changes in wave boundary
conditions. Next we give examples of the natural maximum-entropy principle for
thermal radiation leading to the Planck spectrum, noting that the result
indeed holds for the repulsive case, as was shown earlier for the attractive
case. \ Then we give examples of the attractive and repulsive Casimir forces
due to thermal radiation at finite temperature. In one spatial dimension, we
find that whereas the attractive force increases with increasing temperature,
the repulsive force diminishes to zero with increasing temperature. \ Finally,
we consider the entropy changes of the wave system and note the
temperature-independent Casimir entropy which arises in the attractive but not
the repulsive case at high temperature. \ Some of these aspects of Casimir
forces already appear in the literature for more complicated systems demanding
more elaborate mathematical calculation. \ The examples illustrate the variety
of behaviors encountered for different boundary conditions.

\section{Casimir Forces and Boundary Conditions}

Casimir forces are those arising from the discrete normal mode structure of a
wave system. \ Thus if we consider random radiation in a box, the introduction
of a partition will change the normal mode structure of the box and so lead to
changes in forces and energies. \ The changes of these forces and energies
associated with the placement of the partition are termed Casimir energies.
\ The nature of these Casimir forces and energies will depend upon the energy
spectrum of the waves, upon the number of spatial dimensions, and upon the
boundary conditions at the partition and walls of the box. \ 

In this article we will consider thermodynamic Casimir forces in one space
dimension. \ Thus we may imagine waves on a string between two supports, or
electromagnetic waves all of whose wave vectors point in a single direction
perpendicular to the cavity walls. \ Furthermore, the energy spectrum of the
radiation is not taken arbitrarily[9] but rather is assumed to correspond to
thermodynamic equilibrium and so to satisfy the Wien displacement theorem.
\ Accordingly, the average energy $\mathcal{U}(\omega,T)$ per normal mode is
related to the angular frequency $\omega$ and temperature $T$ by[10]
\begin{equation}
\mathcal{U}(\omega,T)=-\omega\phi^{\prime}(\omega/T)
\end{equation}
where the thermodynamic potential[11] $\phi(\omega/T)\,$ is a function of the
single variable $\omega/T$. \ Of course, the analysis given here can be
extended easily to more general energy spectra.

Wien's displacement theorem for thermal radiation allows two extremes where
the energy $\mathcal{U}$ per normal mode is independent of one of its two
variables. \ If the derivative of the thermodynamic potential is a constant
$\phi^{\prime}(\omega/T)=-const$, then $\mathcal{U}(\omega,T)=const\times
\omega$ corresponding to the familiar zero-point energy where the constant is
chosen as related to Planck's constant,
\begin{equation}
\mathcal{U}=(1/2)\hbar\omega
\end{equation}
\ On the other hand, if the derivative of the thermodynamic potential is equal
to the inverse of its argument $\phi^{\prime}(\omega/T)=-const\times
(T/\omega)$, then we find the familiar energy equipartition result
$\mathcal{U}=const\times T$ where the constant is chosen as Boltzmann's
constant,
\begin{equation}
\mathcal{U}=k_{B}T
\end{equation}

\subsection{Casimir Energy for Zero-Point Radiation\ }

At zero temperature $T=0$, a one-dimensional box of length $L$ contains the
zero-point energy
\begin{equation}
U_{zp}(L)=\Sigma_{n}(1/2)\hbar\omega_{n}%
\end{equation}
where the sum is over all normal modes of wave oscillation. \ If a partition
is introduced at $x$ into the box of length $L$, then the Casimir energy is
the change of energy $\Delta U_{zp}(x,L)$ when the partition is at distance
$x$ from one of the ends compared to when it is in the middle of the box is
\begin{equation}
\Delta U_{zp}(x,L)=U_{zp}(x)+U_{zp}(L-x)-[U_{zp}(L/2)+U_{zp}(L/2)]
\end{equation}
where we have summed over all the energies on both sides of the partition for
the two cases. \ Realistic systems will not enforce the boundary conditions at
very small wavelengths so that these sums should actually involve only a
finite number of waves. \ It is natural to imagine a smooth cut-off function
$F$ dependent upon wavelength $\Lambda$ of the form $F(\Lambda\omega/c)$ where
$c$ is the wave speed. \ One example is $F(\Lambda\omega/c)=\exp
(-\Lambda\omega/c)$. \ The energy sums in (5)\ are evaluated for a finite
cut-off $\Lambda$, and then the cut-off is taken to zero to remove the
dependence upon the cut-off. \ It turns out that the energy found in this
limit is independent of the the choice of the smooth cut-off function
$F$.[12]\ Provided that the boundary conditions are enforced down to some
wavelength $\Lambda$ smaller than any of the dimensions of the system, the
cut-off result for finite $\Lambda$ will agree with the no-cut-off limit
$\Lambda=0$.

For Casimir forces in nature due to electromagnetic fields where we wish to
avoid the details of the interaction with the walls, it is natural to choose
either perfectly-conducting boundary conditions or else infinitely-permeable
boundary conditions.[13] \ In one spatial dimension, these correspond to
Dirichlet or Neumann boundary conditions analogous to a fixed or free end of a
string.[14] \ If we imagine the partition to enforce the same boundary
conditions as the walls (e.g., both partition and walls are good conductors
$\widetilde{=}$ fixed ends), then the normal modes on a string of length $L$
will have wave numbers $k_{n}=n\pi/L$, $n=1,2,3,....$ \ On the other hand, if
the partition enforces the opposite boundary conditions from the walls (e.g.,
a highly permeable partition and good conducting walls $\widetilde{=}$ free
end and fixed end), \ then the normal modes on a string of length $L$ have
wave numbers $k_{n}=(n-1/2)\pi/L$, $n=1,2,3,...$. \ We can discuss both cases
at once by writing $k_{n}=(n-\alpha/2)\pi/L$ where $\alpha=0$ for like
boundary conditions for partition and walls, and $\alpha=1$ for unlike
boundary conditions.

The Casimir energy at zero-temperature can be calculated easily by using an
exponential cut-off function and summing the geometric series,%

\[
\Delta U_{zp}(x,L)=lim_{\Lambda\rightarrow0}\left\{
{\displaystyle\sum\limits_{n=1}^{\infty}}
\frac{\hbar}{2}\frac{c\pi(n-a/2)}{x}\exp\left(  -\Lambda\frac{\pi(n-a/2)}%
{x}\right)  +\right.
\]%
\[
\left.  +%
{\displaystyle\sum\limits_{n=1}^{\infty}}
\frac{\hbar}{2}\frac{c\pi(n-a/2)}{L-x}\exp\left(  -\Lambda\frac{\pi
(n-a/2)}{L-x}\right)  -2%
{\displaystyle\sum\limits_{n=1}^{\infty}}
\frac{\hbar}{2}\frac{c\pi(n-a/2)}{L/2}\exp\left(  -\Lambda\frac{\pi
(n-a/2)}{L/2}\right)  \right\}
\]%
\[
=lim_{\Lambda\rightarrow0}\left\{  -\frac{\hbar c}{2}\frac{\partial}%
{\partial\Lambda}\left[  \frac{\exp(\frac{\Lambda\pi(1-a/2)}{x})}%
{1-\exp(-\frac{\Lambda\pi}{x})}+\frac{\exp(\frac{\Lambda\pi(1-a/2)}{L-x}%
)}{1-\exp(-\frac{\Lambda\pi}{L-x})}-2\frac{\exp(\frac{\Lambda\pi(1-a/2)}%
{L/2})}{1-\exp(-\frac{\Lambda\pi}{L/2})}\right]  \right\}
\]%
\[
=lim_{\Lambda\rightarrow0}\left\{  \left[  \frac{\hbar cx}{\Lambda^{2}\pi
}-\left(  \frac{1}{24}-\frac{\alpha}{8}+\frac{\alpha^{2}}{16}\right)
\frac{c\pi}{x}+\bigcirc(\Lambda)\right]  +\right.
\]%
\[
\left.  +\left[  \frac{\hbar c(L-x)}{2\Lambda^{2}\pi}-\left(  \frac{1}%
{24}-\frac{\alpha}{8}+\frac{\alpha^{2}}{16}\right)  \frac{c\pi}{(L-x)}%
+\bigcirc(\Lambda)\right]  \right.
\]%
\[
\left.  -2\left[  \frac{\hbar c(L/2)}{2\Lambda^{2}\pi}-\left(  \frac{1}%
{24}-\frac{\alpha}{8}+\frac{\alpha^{2}}{16}\right)  \frac{c\pi}{(L/2)}%
+\bigcirc(\Lambda)\right]  \right\}
\]%
\begin{equation}
=-\pi\hbar c\left(  \frac{1}{24}-\frac{\alpha}{8}+\frac{\alpha^{2}}%
{16}\right)  \left(  \frac{1}{x}+\frac{1}{L-x}-\frac{2}{L/2}\right)
\end{equation}
Thus for partition and walls which enforce like boundary conditions (either
both Dirichlet or both Neumann), the parameter $\alpha$ takes the value
$\alpha=0$, and the Casimir energy at zero temperature is negative
\begin{equation}
\Delta U_{zp}(x,L)=-\frac{\pi\hbar c}{24}\left(  \frac{1}{x}+\frac{1}%
{L-x}-\frac{2}{L/2}\right)
\end{equation}
giving an attractive force. \ On the other hand, for partition and walls which
enforce unlike boundary conditions, one Dirichlet and one Neumann, $\alpha=1$
and the Casimir energy is positive
\begin{equation}
\Delta U_{zp}(x,L)=\frac{\pi\hbar c}{48}\left(  \frac{1}{x}+\frac{1}%
{L-x}-\frac{2}{L/2}\right)
\end{equation}
giving a repulsive force. \ 

Exactly the analogous situation is found for the electromagnetic Casimir
forces between two parallel conducting plates or between two permeable plates,
or between a conducting plate and a permeable plate. \ The original 1948
calculation by Casimir[2] involved two conducting plates, and the result has
been recalculated from many points of view.[1] \ The force between a
conducting plate and a permeable plate was given[13] in 1974 and has been
recalculated several times recently.[15]

\subsection{\textbf{CASIMIR ENERGY FOR THE RAYLEIGH-JEANS SPECTRUM}}

The zero-point-energy limit of Wien's law leads to the Casimir energies given
in (7) and (8). \ It is also of interest to obtain the Casimir energies for
the energy-equipartition limit of Wien's law. \ This corresponds to the
Rayleigh-Jeans spectrum of radiation. \ The Casimir energies again take the
form given in (5), but this time with the energy per normal mode in (3).
\ Once again the calculation can easily be carried through analytically for
both like and unlike boundary conditions,
\[
\Delta U_{RJ}(x,L,T)=
\]%
\[
=lim_{\Lambda\rightarrow0}\left\{
{\displaystyle\sum\limits_{n=1}^{\infty}}
k_{B}T\exp\left(  -\Lambda\frac{\pi(n-\alpha/2)}{x}\right)  +%
{\displaystyle\sum\limits_{n=1}^{\infty}}
k_{B}T\exp\left(  -\Lambda\frac{\pi(n-\alpha/2)}{L-x}\right)  +\right.
\]%
\[
\left.  -2%
{\displaystyle\sum\limits_{n=1}^{\infty}}
k_{B}T\exp\left(  -\Lambda\frac{\pi(n-\alpha/2)}{L/2}\right)  \right\}
\]%
\[
=lim_{\Lambda\rightarrow0}\left\{  \frac{k_{B}T\exp(\frac{-\Lambda\pi
(1-\alpha/2)}{x})}{1-\exp(\frac{-\Lambda\pi}{x})}+\frac{k_{B}T\exp
(\frac{-\Lambda\pi(1-\alpha/2)}{L-x})}{1-\exp(\frac{-\Lambda\pi}{L-x})}%
-2\frac{k_{B}T\exp(\frac{-\Lambda\pi(1-\alpha/2)}{L/2})}{1-\exp(\frac
{-\Lambda\pi}{L/2})}\right\}
\]%
\[
=lim_{\Lambda\rightarrow0}\left\{  k_{B}T\left[  \frac{x}{\Lambda\pi}+\frac
{1}{2}(\alpha-1)+\frac{\pi\Lambda}{x}C-...\right]  +k_{B}T\left[  \frac
{L-x}{\Lambda\pi}+\frac{1}{2}(\alpha-1)+\frac{\pi\Lambda}{(L-x)}C-...\right]
+\right.
\]%
\begin{equation}
\left.  -2k_{B}T\left[  \frac{L/2}{\Lambda\pi}+\frac{1}{2}(\alpha-1)+\frac
{\pi\Lambda}{(L/2)}C-...\right]  \right\}  =0
\end{equation}
In both cases, like or unlike boundary conditions, the Rayleigh-Jeans spectrum
gives no change in the Casimir energy. Indeed, the equipartition
Rayleigh-Jeans spectrum is the unique spectrum which produces no Casimir
energy changes associated with the placement of the Casimir partition, $\Delta
U_{RJ}(x,L,T)=0$, no matter whether the partition enforces the same boundary
conditions as the walls or the opposite boundary conditions.

The vanishing changes in Casimir energies for the Rayleigh-Jeans spectrum
found here in (9) agrees with the work of Revzen, Opher, Opher, and Mann[16]
for the case of two parallel conducting plates and with the work of da Silva,
Matos Neto, Placido, Revzen, and Santana[17] for the case of two parallel
plates, one of which is conducting and one of which is permeable.

\section{\noindent Natural Maximum Entropy Principle for Thermal Radiation}

It has been pointed out recently that Casimir forces and zero-point radiation
allow the derivation of the Planck spectrum as the unique radiation spectrum
giving minimum Casimir energy changes over an ensemble of containers with
partitions spaced uniformly across the volume.[18] \ The result has been
submitted for publication in the case of one-dimensional Casimir energies and
has been suggested in the electromagnetic case for a three-dimensional box,
both cases assuming Dirichlet boundary conditions. \ The same energies are
involved for Neumann boundary conditions. \ Here we note that the result holds
also for mixed boundary conditions in the case of one-dimensional waves.

We consider an ensemble of one-dimensional boxes of length $L$ with partitions
placed at different locations $x$ in the boxes so as to give uniform spacing
across the entire ensemble. \ If each of these boxes contains radiation at
thermal equilibrium at the same temperature $T$, then the boxes will still
have different average energies corresponding to different Casimir energies
relating to the differing placements of their partitions. \ Just as it is
natural to suggest that maximum particle entropy involves a uniform
distribution of particles across the volume of a box, it is natural to suggest
that maximum radiation entropy involves the least possible variation in
Casimir energies across the ensemble of boxes. \ Thus we are looking for that
spectrum of random radiation which satisfies the Wien displacement theorem
(and therefore has energy per normal mode $k_{B}T$ per normal mode at low
frequency and $(1/2)\hbar\omega$ at high frequency) which gives a minimum for
the sum of the absolute values of the Casimir energies taken across the
ensemble of boxes,%

\begin{equation}
I=\Sigma_{i}|\Delta U(x_{i},L,T)|
\end{equation}
\ This condition can be converted to the minimum for an integral%
\begin{equation}
I=%
{\displaystyle\int\limits_{x=\delta}^{x=L/2}}
dx\left\vert \Delta U(x,L,T)\right\vert
\end{equation}
where $\delta$ is a fixed small cut-off distance. \ The fixed distance
$\delta$ is chosen much smaller than any distance of interest in the
situation; it is required by the divergence of the Casimir energy at small
separations. \ 

The integrand in (11) can be evaluated numerically for various suggested
radiation spectra which provide smooth interpolations between $k_{B}T$ at low
frequency and $(1/2)\hbar\omega$ at high frequency. \ Indeed one can assume
various functional forms and then vary the parameters to provide the smallest
value for the integral (11). \ For example, if one introduces parameters
$C_{1}$ and $C_{2}$ into the functional form
\begin{equation}
\mathcal{U}_{C_{1}C_{2}}(\omega,T)=\frac{C_{1}\omega\exp[-C_{2}(\omega
/T)]}{1-\exp[-C_{1}(\omega/T)]}+(1/2)\omega
\end{equation}
then this provides a smooth interpolation of the required form for every
positive value of $C_{1}$ and $C_{2}$. \ By numerical calculation for both
like and unlike boundary conditions, it is found that the values of $C_{1}$
and $C_{2}$ which provide the smallest integral $I$ in (11) are $C_{1}=1$ and
$C_{2}=1$, corresponding to exactly the Planck spectrum with zero-point energy%
\begin{equation}
\mathcal{U}_{Pzp}\left(  \omega,T\right)  =\frac{\omega\exp[-(\omega
/T)]}{1-\exp[-(\omega/T)]}+\frac{1}{2}\omega=\frac{\omega}{\exp(\omega
/T)-1}+\frac{1}{2}\omega=\frac{1}{2}\omega\coth\left(  \frac{1}{2}\frac
{\omega}{T}\right)
\end{equation}
where for the numerical evaluation we have set $\hbar/k_{B}=1$. The case of
like boundary conditions where the normal modes of the wave field are
$k_{n}=n\pi/L$, $n=1,2,....$ was calculated earlier and submitted for
publication.[18] \ The case of unlike boundary conditions where the normal
modes of the field are $k_{n}=(n-1/2)\pi/L$ has now been completed and is
found by numerical calculation to lead to the Planck law as the spectrum
providing a minimum for the integral $I$ in (11). \ For this case and for all
spectra which have been tested, the Planck spectrum with zero-point radiation
gave a smaller integral in (11).

An idea of what is involved can be seen in the Figs 1, 2, and 3. \ In Fig. 1,
we plot the energy per normal mode $\mathcal{U}$ versus frequency $\omega$ for
zero-point radiation, for the Rayleigh-Jeans spectrum, and for the Planck
spectrum with zero-point radiation at two different temperatures. \ We note
how smoothly the Planck spectrum interpolates between the equipartition
behavior at low frequency and zero-point energy at high frequency. \ Figures 2
and 3 give the Casimir energies in a one-dimensional box with a partition at
location $x$ for these same radiation spectra. \ In Figure 2 (like boundary
conditions), the partition has the same boundary conditions as the walls of
the box, while in Figure 3 (unlike boundary conditions), the partition has
opposite boundary conditions from the walls. Thus we see the negative
zero-point Casimir energy (7) in Fig. 2 and the positive zero-point Casimir
energy (8) in Figure 3. \ The Rayleigh-Jeans spectrum gives zero Casimir
energies (9) for both cases. The Planck spectrum with zero-point energy
provides a smooth interpolating Casimir energy between these two extremes.
\ One should notice how the Planck spectrum with zero-point energy hugs the
axis of both the figures 3 and 4 before leaving the axis to join the
zero-point curve at small separations $x$. \ This hugging-the-axis corresponds
to small Casimir energies and accounts for the minimum behavior of the test
integral in (11). \ It is easy to graph the Casimir energies for other
interpolating spectra. \ One finds that other interpolating spectra satisfying
the Wien displacement law do not hug the axis nearly so closely and therefore
do not give a minimum for the test integral (11).[19]

We note that in both these cases, the presence of zero-point radiation is
absolutely necessary. \ In the absence of zero-point radiation, there would be
no non-zero minimum for the integral $I$; rather the energy would diverge to
the Rayleigh-Jeans spectrum where the Casimir energies vanish for both like
and unlike boundary conditions. \ It is precisely the presence of zero-point
radiation which prevents the familiar "ultraviolet catastrophe" for thermal
radiation. \ 

\section{Casimir Forces at Finite Temperature}

At zero temperature, the Casimir forces $X_{zp}$ on a partition can be
obtained by using the Casimir energies in (7) and (8) as potential functions%
\begin{equation}
\Delta X_{zp}=-d(\Delta U_{zp})/dx
\end{equation}
Accordingly, we find the attractive force
\begin{equation}
\Delta X_{zp}=-\frac{\pi\hbar c}{24}\left(  \frac{1}{x^{2}}-\frac{1}%
{(L-x)^{2}}\right)
\end{equation}
for the case of like boundary conditions ($\alpha=0$), and the repulsive force%
\begin{equation}
\Delta X_{zp}=\frac{\pi\hbar c}{48}\left(  \frac{1}{x^{2}}-\frac{1}{(L-x)^{2}%
}\right)
\end{equation}
for the case of unlike boundary conditions ($\alpha=1$).

At finite temperature $T$, it is the derivative of the change in the Helmholtz
free energy $\Delta F$ which gives the force on the partition. \ The Helmholtz
free energy $\mathcal{F}(\omega,T)$ per wave normal mode in the case of the
Planck spectrum with zero-point radiation is given by
\begin{equation}
\mathcal{F}_{Pzp}(\omega,T)=k_{B}T\ln\left[  2\sinh\left(  \frac{\hbar\omega
}{2k_{B}T}\right)  \right]  =\frac{1}{2}\hbar\omega+k_{B}T\ln\left[
1-\exp\left(  \frac{-\hbar\omega}{k_{B}T}\right)  \right]
\end{equation}
The Helmholtz free energy in a box is found by summing the Helmholtz free
energies of the radiation modes
\begin{equation}
F(L,T)=\Sigma_{n}\mathcal{F}(\omega_{n},T)
\end{equation}
and the change in the Helmholtz free energy due to a change in the position of
a partition in the box is
\begin{equation}
\Delta F(x,L,T)=F(x,T)+F(L-x,T)-2F(L/2,T)
\end{equation}
Because of the zero-point energy, the change in Helmholtz free energy for the
Planck spectrum with zero-point radiation requires the use of a temporary cut
off as for our calculation in Eq. (6). \ For numerical evaluation, one may
separate out the divergent zero-point energy contribution $\Delta F_{zp}$ and
deal with the convergent thermal part $\Delta F_{PT}=\Delta F_{Pzp}-\Delta
F_{zp}$. \ The zero-point Helmholtz free energy $\Delta F_{zp}$ is identical
to the zero-point energy $\Delta U_{zp}$ given in (6). \ 

However, rather than dealing with the total Helmholtz free energy $\Delta
F(x,L,T)$ as a potential function for forces, it seems easier to sum the
forces due to the individual normal modes. \ The force $\mathcal{X}%
(\omega,L,T)$ on the boundary due to a single normal mode in a box is given by
the derivative of the Helmholtz free energy of the normal mode in (17)%
\begin{equation}
\mathcal{X}_{Pzp}(\omega,L,T)=-\left(  \frac{\partial\mathcal{F}_{Pzp}%
(\omega,T)}{\partial\omega}\right)  _{T}\frac{d\omega}{dL}=\frac{\hbar\omega
}{2L}\coth\left(  \frac{\hbar\omega}{2k_{B}T}\right)
\end{equation}
Then the force $\Delta X$ on a partition at $x$ due to all the modes is
\begin{equation}
\Delta X(x,L,T)=%
{\displaystyle\sum\limits_{n=1}^{\infty}}
\left[  \mathcal{X}(\omega_{n},x,T)-\mathcal{X}(\omega_{n},L-x,T)\right]
\end{equation}
where once again we can include both sets of boundary conditions by including
the parameter $\alpha$ in the normal modes in (21). \ For the Planck spectrum
with zero-point radiation used in (20), this sum can be easily evaluated in
closed form for the limits of zero-point radiation (which gives just $\Delta
X_{Pzp}$ above in (15) and (16)) or for the high-temperature limit of the
Rayleigh-Jeans spectrum. \ For finite temperatures, the force can easily be
evaluated numerically; we separate off the zero-point contribution and are
left with a convergent sum in (21). \ 

In the case of the high-temperature equipartition limit corresponding to the
Rayleigh-Jeans law, the force in (20) due to a single normal mode becomes
independent of $\omega$,%
\begin{equation}
\mathcal{X}_{RJ}(\omega,L,T)=\frac{k_{B}T}{L}%
\end{equation}
Therefore the sum over all the normal modes in (21) now involves the
difference of two divergent sums. \ Accordingly, we temporarily introduce a
smooth frequency-dependent cut-off function dependent upon the cut-off
parameter $\Lambda$, and take the net force as the result when $\Lambda
\rightarrow0$,%
\[
\Delta X_{RJ}(x,L,T)=lim_{\Lambda\rightarrow0}\left\{
{\displaystyle\sum\limits_{n=1}^{\infty}}
\frac{k_{B}T}{x}\exp\left(  -\Lambda\frac{c\pi(n-\alpha/2)}{x}\right)
+\right.
\]

\begin{equation}
\left.  -%
{\displaystyle\sum\limits_{n=1}^{\infty}}
\frac{k_{B}T}{L-x}\exp\left(  -\Lambda\frac{c\pi(n-\alpha/2)}{L-x}\right)
\right\}
\end{equation}
Once again we are dealing with a geometric series which can be summed easily
to give%
\[
\Delta X_{RJ}(x,L,T)=lim_{\Lambda\rightarrow0}\left\{  \frac{k_{B}T}{x}\left[
\frac{x}{\Lambda\pi}+\frac{1}{2}\left(  \alpha-1\right)  +\frac{\Lambda\pi}%
{x}C+...\right]  +\right.
\]

\[
\left.  -\frac{k_{B}T}{L-x}\left[  \frac{L-x}{\Lambda\pi}+\frac{1}{2}\left(
\alpha-1\right)  +\frac{\Lambda\pi}{L-x}C+...\right]  \right\}
\]%
\begin{equation}
=\frac{k_{B}T}{2}\left(  \alpha-1\right)  \left(  \frac{1}{x}-\frac{1}%
{L-x}\right)
\end{equation}
Thus at high temperatures for like boundary conditions where $\alpha=0$, there
is a force attracting a partition to one of the walls
\begin{equation}
\Delta X_{RJ}(x,L,T)=-\frac{k_{B}T}{2}\left(  \frac{1}{x}-\frac{1}%
{L-x}\right)
\end{equation}
We notice that this force, which arises from the discrete classical normal
mode structure of the box, involves a different power of length from the
zero-point force in (15), but both are significant only if $x$ or $L-x$ is
small. \ On the other hand, in one dimension for unlike boundary conditions
where $\alpha=1$, the net force on the partition in (24) vanishes entirely
$\Delta X_{RJ}(x,L,T)=0$ in the high-temperature Rayleigh-Jeans limit. \ \ 

The results for the Casimir forces on a partition in one space dimension are
shown in Figures 4 and 5 for like and unlike boundary conditions respectively.
\ In the case of like boundary conditions shown in Fig. 4, the Casimir force
is attractive at all separations and increases steadily as the temperature
increases. \ In the case of unlike boundary conditions shown in Fig. 5, the
Casimir force is repulsive at low temperatures for fixed displacement $x$, but
then goes to zero as the temperature increases.\ In the limit of high
temperature, both force curves go over to the results given by the
Rayleigh-Jeans limit.

We should emphasize that the appearance of attractive or repulsive Casimir
forces at zero temperature is not intuitively clear; changes in boundary
conditions or shape or spatial dimension all seem involved.[1] \ Also, there
seems to be some disagreement regarding the detailed temperature dependence of
Casimir forces for like boundary conditions, though only increasingly
attractive forces with increasing temperature seem to be expected, just as
seen in Fig. 4.[20] \ In the case of repulsive forces involving mixed boundary
conditions, the one-dimensional result in Fig. 5 involving diminishing
repulsion with increasing temperature agrees with the behavior found by Ferrer
and Grifols[21] for the repulsive Casimir-Polder force involving the electric
polarizability of one particle and the magnetic polarizability of another.[22]
\ However, it is qualitatively different from the increasing repulsion with
temperature found by da Silva, Matos Neto, Placido, Revzen, and Santana for
the high-temperature limit of two parallel plates one of which is a conductor
and the other of which is highly permeable.[17]

\section{Casimir Entropy}

In addition to the surprising sign reversals found for Casimir forces and
energies, and to the natural connection of Casimir energies with the Planck
spectrum, there is an entropy aspect of the Casimir forces which seems
curious. \ The Casimir entropy is the change in entropy of a wave system
depending upon the placement of the system partition. The entropy $S(L,T)$ of
radiation in a box is the sum over modes for the entropy \ $\mathcal{S}%
(\omega/T)$\ of a normal mode which follows from the thermodynamic potential
$\phi(\omega/T)$ as%
\begin{equation}
\mathcal{S}(\omega/T)=\phi(\omega/T)-\frac{\omega}{T}\phi^{\prime}%
(\omega/T)\text{ \ \ \ \ and \ \ \ }S(L,T)=%
{\displaystyle\sum\limits_{n=1}^{\infty}}
\mathcal{S}(\omega_{n}/T)
\end{equation}
Accordingly, the Casimir entropy change $\Delta S(x,L,T)$ in a box of length
$L$ at temperature $T$ due to a partition at $x$, is given by
\begin{equation}
\Delta S(x,L,T)=S(x,T)+S(L-x,T)-2S(L/2,T)
\end{equation}
For the Planck spectrum%
\begin{equation}
\mathcal{S}_{P}\left(  \frac{\omega}{T}\right)  =-\ln\left[  2\sinh\left(
\frac{\hbar\omega}{2k_{B}T}\right)  \right]  +\frac{\hbar\omega}{2k_{B}T}%
\coth\left(  \frac{\hbar\omega}{2k_{B}T}\right)
\end{equation}
the entropy changes for the case of like boundary conditions are given in Fig.
6, and those for unlike boundary conditions are given in Fig. 7. \ It is
curious that in the case of like boundary conditions shown in Fig. 6, there is
clearly a non-zero, temperature-independent limit for the Casimir entropy
$\Delta S(x,L,T)$ at high temperature, whereas in the case of unlike boundary
conditons shown in Fig. 7 there is not.

\subsection{\noindent\textbf{ENTROPY CHANGES IN THE HIGH-TEMPERATURE LIMIT}}

For fixed frequency $\omega$ and increasing temperature $T$, the Planck
spectrum with zero-point radiation goes over to the Rayleigh-Jeans spectrum.
Similarly, for fixed length $x$ and increasing temperature $T$ the Casimir
energy $\Delta U_{Pzp}(x,L,T)$ for the Planck spectrum with zero-point
radiation goes over to the Rayleigh-Jeans result $\Delta U_{RJ}(x,L,T)=0$.
However, the net force $\Delta X_{Pzp}(x,L,T)$\ and the net change in entropy
$\Delta S_{P}(x,L,T)$ do not necessarily go to zero with increasing
temperature. \ Indeed, these quantities go over to the values for the
Rayleigh-Jeans spectrum, which values need not vanish. Thus at high
temperature, Casimir forces are associated with changes in system entropy, not
the vanishing changes in system energy.

Applying thermodynamic analysis to a partitioned one-dimensional box of
radiation with a Rayleigh-Jeans spectrum at temperature $T,$ an isothermal
change of the the position of the partition gives no change in system energy
but involves heat added so that%

\begin{equation}
Td\Delta S_{RJ}=d\Delta U_{RJ}+\Delta X_{RJ}dx=0+\Delta X_{RJ}dx
\end{equation}
Then using the Casimir force (25) associated with the Rayleigh-Jeans spectrum
and like boundary conditions,%

\begin{equation}
d\Delta S_{RJ}=\frac{X_{RJ}}{T}dx=\frac{-k_{B}}{2}\left(  \frac{1}{x}-\frac
{1}{L-x}\right)  dx
\end{equation}
and%

\begin{equation}
\Delta S_{RJ}(x,L,T))=\frac{k_{B}}{2}\left\vert \ln\left(  \frac{x}%
{L-x}\right)  \right\vert
\end{equation}
We emphasize that the right-hand side of Eq.(31) is independent of temperature
$T$. Thus due to the normal mode structure of the box, there is a
temperature-independent change of entropy when the partition is moved. This
seems reminiscent of the temperature-independent change associated with mixing
entropy for ideal gas particles. This same result (31) can be obtained by
summing the Rayleigh-Jeans entropy $\mathcal{S}(\omega/T)=-\ln(\omega/T)+1$
per normal mode over the normal modes of the box when using a cut-off
function, subtracting, and allowing the cut-off to disappear. \ 

The Casimir entropy change $\Delta S_{P}(x,L,T)=S_{P}(T,x)+S_{P}%
(T,L)-2S_{P}(T,L/2)$ associated with the Planck spectrum (and involving
convergent sums) goes to the Rayleigh-Jeans limit (31) at high temperature for
fixed values of $x$ and $L$. However, for fixed finite temperature $T$, the
entropy change $\Delta S_{P}(x,L,T)$ for the Planck spectrum goes to a finite
temperature-dependent limit as $x$ goes to $0$ or $L$, as seen in Fig. 6.

The Casimir entropy for unlike boundary conditions is seen in Fig. 7 to first
increase from zero for increasing temperature and then to decrease back to
zero for additional increase of temperature for fixed partition-coordinate x.
\ Thus in one dimension for unlike boundary conditions, there is no
non-vanishing temperature-independent Casimir entropy in the high-temperature
limit. \ We note that for the Rayleigh-Jeans spectrum, there is no Casimir
energy $\Delta U_{RJ}=0$ and, for unlike boundary conditions in one dimension,
no Casimir force $\Delta\dot{X}_{RJ}$. \ Accordingly, the first law of
thermodynamics in (29) indicates that the Casimir entropy vanishes for the
Rayleigh-Jeans spectrum and unlike boundary conditions.

The existence of temperature-independent Casimir entropy seems to depend
sensitively on both the boundary conditions and dimension. \ The existence of
a geometrical temperature-independent Casimir entropy for like boundary
conditions as given in (31) agrees with the result reported by Revzen, Opher,
Opher and Mann for the case of two conducting parallel plates.[23] \ However,
the absence of such an entropy for unlike boundary conditions is in contrast
to the temperature-independent Casimir entropy reported by da Silva, Matos
Neto, Pacido, Revzen, and Santana for the case two parallel plates, one of
which is conducting and one of which is highly permeable.[17]

\section{Closing Summary}

The the introduction of a partition into a wave system leads to changes in
energies, forces, and thermodynamic variables associated with the wave
boundary conditions at the partition and the walls. \ For electromagnetic
waves, these Casimir forces are termed "dispersion forces" or "van der Waals
forces." \ Although physicists usually think of van der Waals forces as
attractive, the type of analysis introduced originally by Casimir suggests the
possibility of repulsive forces. \ Several aspects of Casimir forces can be
easily illustrated by a mathematical model involving waves in one space
dimension. \ Depending upon the boundary conditions, a partition introduced
into a one-dimensional box will experience attraction to or repulsion form the
walls at zero temperature. \ The attractive force becomes increasingly
attractive with increasing temperature whereas the repulsive force diminishes
with increasing temperature. \ Also, there are changes in entropy associated
with changes in position for the partition. \ For high temperatures, these
entropy changes become temperature independent in the attractive case (rather
like a wave form of mixing entropy) but vanish in the repulsive case.
\ Finally, we note that Casimir energies are closely related to the
thermodynamics of blackbody radiation, allowing a derivation of the Planck
spectrum from a natural maximum-entropy principle.

\section{\ Acknowledgement:}

\ I would like to thank Professor Daniel C. Cole for sending me annotated
copies of the work on repulsive Casimir forces by E. Buks and M. L. Roukes,
and by O. Kenneth, I. Klich, A.. Mann, and M. Revzen . \ It was Professor
Cole's communication which stimulated the present work.

\pagebreak

\begin{center}
\textbf{Figure Captions}
\end{center}

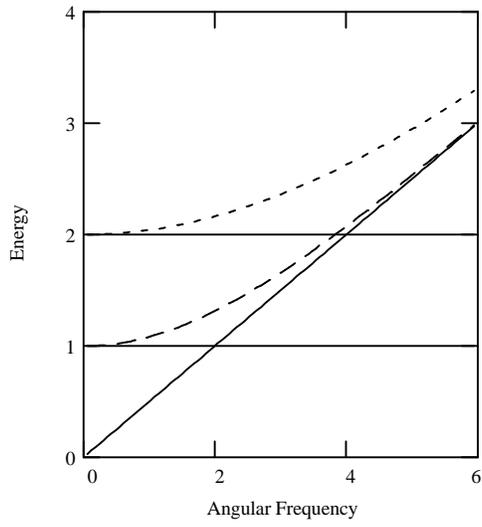
Figure 1: Planck Spectrum Energy Per Normal Mode. \ The energy $\mathcal{U}%
_{Pzp}(\omega,T)$ of the Planck spectrum with zero-point energy (13) is
plotted vs mode angular frequency for three different temperatures. The
sloping solid line for $T=0$ corresponds to the zero-point energy
$\mathcal{U}_{zp}=(1/2)\hbar\omega$. \ The two horizontal lines correspond to
the Rayleigh-Jeans spectra for the temperatures $T=1$ and $T=2$. \ The Planck
expression for the energy per normal mode at $T>0$ starts with the
equipartition value at low frequency and then goes smoothly to the zero-point
curve at high frequency. The dashed curve for $T=1$ starts with the
equipartition value $\mathcal{U}=1$ at low frequency, while the dotted curve
for $T=2$ starts with the equipartition value $\mathcal{U}=2$. \ The Planck
law interpolates smoothly between the Rayleigh-Jeans and zero-point spectra.

Figure 2: Casimir Energy for Like Boundary Conditions. $\ $The Casimir energy
$\Delta U_{Pzp}(x,L,T)$, for the Planck spectrum with zero-point radiation in
the case of like boundary conditions between walls and partition, is plotted
as a function of the distance $x$ from the left-hand wall located at $x=0$ in
a box of length $L=3$ for three different temperatures. \ The solid curve
gives the negative Casimir energy (7) at temperature $T=0$. \ The dashed curve
for $T=1$ gives Casimir energies close to zero near the center of the box
before dropping down near the walls to join the curve for zero-point energy.
\ The dotted curve at $T=3$ hugs the axis for a larger central interval. \ The
Rayleigh-Jeans result (9), which holds for the high-temperature limit of the
Planck spectrum, has zero Casimir energy for every value of $x$.

Figure 3: Casimir Energy for Unlike Boundary Conditions. \ The Casimir energy
$\Delta U_{Pzp}(x,L,T)$, for the Planck spectrum with zero-point radiation in
the case of unlike boundary conditions between walls and partition, is plotted
as a function of the distance $x$ from the left-hand wall located at $x=0$ in
a box of length $L=3$ for three different temperatures. \ The solid curve
gives the positive Casimir energy (8) at temperature $T=0$. \ The dashed curve
for $T=1$ gives Casimir energies close to zero near the center of the box
before rising near the walls to join the curve for zero-point energy. \ The
dotted curve at $T=3$ hugs the axis for a larger central interval. \ The
Rayleigh-Jeans result (9), which holds for the high-temperature limit of the
Planck spectrum, has zero Casimir energy for all values of $x$.

Figure 4: Casimir Force for Like Boundary Conditions. \ The Casimir force
$\Delta X_{Pzp}(x,L,T)$, for the Planck spectrum with zero-point radiation in
the case of like boundary conditions between walls and partition, is plotted
as a function of the distance $x$ from the left-hand wall located at $x=0$ in
a box of length $L=3$ for three different temperatures. \ The solid curve
gives the Casimir force (15) which attracts the partition to the walls at
temperature $T=0$. \ The dashed curve for $T=1$ shows an increased Casimir
force which near the walls joins the curve for the zero-point Casimir force.
\ The dotted curve at $T=3$ shows the continued increase with increasing
temperature. \ The Rayleigh-Jeans result (25), which holds for the
high-temperature limit of the Planck spectrum, has increasing Casimir force
with increasing temperature.

Figure 5: Casimir Force for Unlike Boundary Conditions. \ The Casimir force
$\Delta X_{Pzp}(x,L,T)$, for the Planck spectrum with zero-point radiation in
the case of unlike boundary conditions between walls and partition, is plotted
as a function of the distance $x$ from the left-hand wall located at $x=0$ in
a box of length $L=3$ for three different temperatures. \ The solid curve
gives the Casimir force (16) which repels the partition from the walls at
temperature $T=0$. \ The dashed curve for $T=1$ shows decreasing Casimir force
in the central region of the box before leaving the axis near the walls to
join the curve for the zero-point Casimir force. \ The dotted curve at $T=3$
shows the continued decrease with increasing temperature. \ The Rayleigh-Jeans
result, which holds for the high-temperature limit of the Planck spectrum, has
vanishing Casimir force at all temperatures for the one-dimensional case with
unlike boundary conditions.

Figure 6: Casimir Entropy for Like Boundary Conditions. \ The Casimir entropy
$\Delta S_{P}(x,L,T)$, for the Planck spectrum with zero-point radiation in
the case of like boundary conditions between walls and partition, is plotted
as a function of the distance $x$ from the left-hand wall located at $x=0$ in
a box of length $L=3$ for several different temperatures. \ At zero
temperature the Casimir entropy vanishes. \ The solid curve gives the Casimir
entropy at $T=0.25$. \ The dashed curve for $T=0.5$ shows increasing Casimir
entropy at this higher temperature away from the central region, which is
continued in the dash-dot curve for $T=1$ and the dotted curve for $T=2$. \ In
the central region, the Casimir entropy goes over to a temperature-independent
result at high temperature. \ The Rayleigh-Jeans result, which holds for the
high-temperature limit of the Planck spectrum, involves a
temperature-independent Casimir entropy (31) at all temperatures for a
one-dimensional box with like boundary conditions.

Figure 7: Casimir Entropy for Unlike Boundary Conditions. \ The Casimir
entropy $\Delta S_{P}(x,L,T)$, for the Planck spectrum with zero-point
radiation in the case of unlike boundary conditions between walls and
partition, is plotted as a function of the distance $x$ from the left-hand
wall located at $x=0$ in a box of length $L=3$ for several different
temperatures. \ At zero temperature the Casimir entropy vanishes. \ The solid
curve gives the Casimir entropy at $T=0.1$. \ The dashed curve for $T=0.2$
shows increasing Casimir entropy at this higher temperature away from the
central region. \ However, the dash-dot curve for $T=0.5$ shows a decrease of
the Casimir entropy in the central region of the box and an increase near the
walls. \ The dotted curve for $T=1$ shows the continued decrease in Casimir
entropy in the central region. \ The Rayleigh-Jeans result, which holds for
the high-temperature limit of the Planck spectrum, involves vanishing Casimir
entropy at all temperatures for a one-dimensional box with unlike boundary conditions.

\bigskip

References

\end{document}